\documentclass[a4paper,11pt]{article}
\usepackage{pos}

\usepackage{slashed}
\usepackage{xcolor}
\usepackage{wrapfig}
\usepackage{ulem}

\title{SymEFT predictions for local fermion bilinears}

\author*[a,b,c]{Nikolai Husung}

\affiliation[a]{Physics and Astronomy, University of Southampton,\\
Southampton SO17 1BJ, United Kingdom}

\affiliation[b]{Instituto de Física Teórica UAM-CSIC,\\
C/ Nicolás Cabrera 13-15, Universidad Autónoma de Madrid, Cantoblanco 28049 Madrid, Spain}

\affiliation[c]{Departamento de Física Teórica, Universidad Autónoma de Madrid,\\
Cantoblanco 28049 Madrid, Spain}

\emailAdd{nikolai.husung@uam.es}

\abstract{Beyond spectral quantities, Symanzik Effective Theory (SymEFT) predictions of the asymptotic lattice-spacing dependence require the inclusion of an additional minimal basis of higher-dimensional operators for each local field involved in the matrix element of interest.
Adding the proper bases for fermion bilinears of mass-dimension 3 allows to generalise previous predictions to matrix elements of those bilinears.
These results can be incorporated in ansätze used in continuum extrapolations and should allow improved control of the associated systematic uncertainties.
Potential difficulties and pitfalls are being highlighted.
The current work is limited to the use of Wilson or Ginsparg-Wilson quarks in both sea and valence.}

\FullConference{The 40th International Symposium on Lattice Field Theory (Lattice 2023)\\
July 31st - August 4th, 2023\\
Fermi National Accelerator Laboratory\\}

\def\vecn{\mathbf{0}}

\def\rmd{\mathrm{d}}
\def\tr{\mathop{\mathrm{tr}}}

\def\ord{\mathop{\mathrm{O}}}

\newcommand{\cev}[1]{\overset{\leftarrow}{#1}}

\def\gbar{\bar{g}}
\def\nmin{n_\mathrm{min}}
\def\Gammahat{\hat{\Gamma}}
\def\L{\mathscr{L}}

\def\op{\mathcal{O}}
\def\opE{\op_{\mathcal{E}}}
\def\Qop{Q}
\def\Qbase{\mathcal{Q}}
\def\base{\mathcal{B}}
\def\baseE{\base_{\mathcal{E}}}
\def\Jop{J}
\def\Jbase{\mathcal{J}}

\newcommand{\Pbase}{\big(\mathcal{P}^{kl}\big)}
\newcommand{\A}[1][\mu]{\big(\mathrm{A}_{#1}^{kl}\big)}
\newcommand{\Abase}[1][\mu]{\big(\mathcal{A}_{#1}^{kl}\big)}

\newcommand{\Tbase}[1][\mu\nu]{\big(\mathcal{T}_{#1}^{kl}\big)}

\def\Nf{N_\mathrm{f}}
\def\Nb{N_\mathrm{b}}

\def\MSbar{\overline{\text{MS}}}

\begin{document}
\maketitle
\normalem

\section{Introduction}
Aiming for predictions of continuum physics of Quantum Chromodynamics (QCD) from lattice QCD simulations, the lattice artifacts from the discretised lattice QCD action and any discretised local fields must eventually be removed to reach the continuum limit.
The continuum extrapolation towards zero lattice spacing $a\searrow 0$ is a common source of systematic uncertainties in our field and must be kept in check.
Although lattice QCD is not a classical field theory the commonly used ansatz for continuum extrapolations is a plain integer-power in the lattice spacing $a^{\nmin}$, typically $\nmin\in\{1,2\}$.
While the overall polynomial lattice-spacing dependence is correct, quantum effects will shift the asymptotically leading dependence towards $a^{\nmin}[2b_0\gbar^2(1/a)]^{\Gammahat_i}$, where $\gbar(1/a)$ is the running coupling going asymptotically to zero.
Here $b_0$ is the 1-loop coefficient of the QCD $\beta$-function and the $\Gammahat_i$ depend on the particular choice for the lattice discretisation of the action and any local fields present in the observable of interest.
As shown in the seminal work of Balog, Niedermayer, and Weisz on the O(3) model \cite{Balog:2009yj,Balog:2009np}, those powers $\Gammahat_i$ can be severely negative (there $\min_i\Gammahat_i=-3$) and therefore spoil the convergence towards the continuum limit.
\begin{wrapfigure}{r}{0.44\textwidth}
\includegraphics[scale=0.95,page=1]{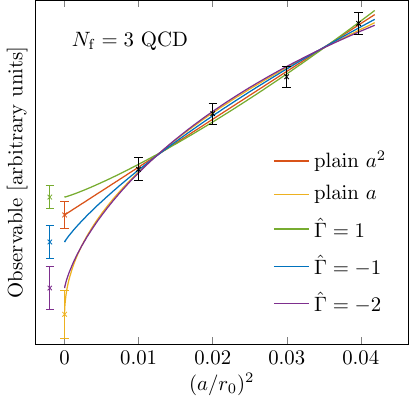}
\caption{Continuum extrapolation of synthetic data points randomly generated according to $a^2[2b_0\gbar^2(1/a)]^{-2}$.
The various ans\"atze extrapolating to a constant include the typical choice of simple $a^2$ corrections and unconventional powers in the coupling $\hat{\Gamma}\in\{-2,-1,1\}$ modifying $a^2[2b_0\gbar^2(1/a)]^{\hat{\Gamma}}$, as well as simple $a$ corrections.
Both $a$ and (the correct ansatz) $a^2[2b_0\gbar^2(1/a)]^{-2}$ agree within uncertainties, while the simple $a^2$ by construction completely misses the curvature.
Three continuum extrapolated values have been shifted to the left for better readability.\\[-3pt]}
\label{fig:contLimitSketch}
\end{wrapfigure}
Not accounting for these effects may then lead to severely underestimated uncertainties or even invalidate the continuum values obtained as sketched in fig.~\ref{fig:contLimitSketch}.
Notice that fig.~\ref{fig:contLimitSketch} is an oversimplified example assuming $a^2[2b_0\gbar^2(1/a)]^{\Gammahat}$ with only one power $\Gammahat=-2$ rather than a full set of those powers combined with subleading corrections suppressed both in the coupling and in the lattice spacing.
In practice the situation will be even more complicated.
We foresee that accounting for these type of systematic uncertainties will play an essential role to consolidate the continuum-limit lattice QCD results, which are relevant inputs for precision physics experiments like e.g.~at the LHC \cite{Pasztor:2019rqu,LHCb:2008vvz}, SuperKEKB \cite{Belle-II:2018jsg}, or the Fermilab muon g-2 experiment~\cite{Muong-2:2023cdq}.

In previous works the leading asymptotic lattice-spacing dependence has been computed for lattice actions in pure gauge \cite{Husung:2019ytz} and full QCD with Wilson or Ginsparg-Wilson (GW) quarks \cite{Husung:2021mfl,Husung:2022kvi} yielding the leading powers $\Gammahat_i^\base$ associated to the discretised action.
While those results are sufficient for spectral quantities like hadron masses, any matrix elements with the insertion of a local field $J(x)$ will have additional powers $\Gammahat_i^J$ originating from the discretised local field.
As a next step, fermion bilinears of mass-dimension~3 are being considered here.
Those are the most commonly used local fields as they are needed for the computation of, e.g., decay constants and form factors.
For a comprehensive overview of lattice QCD determinations of such quantities see FLAG~\cite{FlavourLatticeAveragingGroupFLAG:2021npn}.

\section{Symanzik Effective Field Theory}
If the lattice spacings are sufficiently small, lattice artifacts can be described as a perturbation around continuum QCD in Symanzik's Effective Field Theory (SymEFT) \cite{Symanzik:1979ph,Symanzik:1981hc,Symanzik:1983dc,Symanzik:1983gh}, i.e., we can write the lattice spacing dependence in terms of an effective continuum Lagrangian
\begin{equation}
\L_\mathrm{eff}(x)=\L_\mathrm{QCD}(x)+a^{\nmin}\sum_i\omega_i\Qop_i^{(\nmin)}(x)+\ord(a^{\nmin+1})\,.
\end{equation}
The Lagrangian of continuum QCD involving $\Nf$ sea quarks and $\Nb$ valence quarks\footnote{We implement valence quarks via the addition of both a quark field $\psi_{\Nf+i}$ and a complementing ghost field $\phi_i$ with $i\in\{1,...,\Nb\}$ to cancel any contribution to the sea, see e.g.~\cite{Bar:2005tu}, thus allowing for mixed actions and optionally some quenched flavours.
More complicated choices for the sea or valence sector are possible but enlarge the operator bases further.} reads
\begin{align}
\L_\mathrm{QCD}&=-\frac{1}{2g_0^2}\tr(F_{\mu\nu}F_{\mu\nu})+\bar{\Psi}\left\{\gamma_\mu D_\mu(A)+M\right\}\Psi\,,
&\Psi^T&=(\psi_1,...,\psi_{\Nf+\Nb},\phi_1,...,\phi_{\Nb})\,,
\end{align}
where $F_{\mu\nu}=[D_\mu(A),D_\nu(A)]$ is the field strength tensor, $D_\mu(A)=\partial_\mu+A_\mu$ is the covariant derivative, $A_\mu\in\mathrm{su}(N)$, and $\omega_i$ are (bare) matching coefficients.
The lattice artifacts are described by the higher-dimensional operators $\Qop_i^{(\nmin)}$ of mass-dimension $(4+\nmin)$, which form a minimal basis compatible with the symmetries realised by the chosen lattice action.
Here we restrict considerations to Wilson quarks~\cite{Wilson:1974,Wilson:1975id} or GW quarks~\cite{Ginsparg:1981bj}, i.e., (graded) flavour symmetries $\mathrm{SU}(\Nf)_\mathrm{V}\times U(1)_\mathrm{B}$ or $\mathrm{SU}(\Nf)_\mathrm{L}\times\mathrm{SU}(\Nf)_\mathrm{R}\times U(1)_\mathrm{B}$ respectively in the sea (valence).
For a SymEFT analysis of such optionally mixed actions see \cite{Husung:2022kvi}.
In contrast to this earlier work we now need to keep track of operators vanishing by the classical EOMs
\begin{equation}
[D_\mu(A), F_{\mu\nu}]=T^ag_0^2\bar\Psi\gamma_\nu T^a\Psi,\quad \gamma_\mu D_\mu(A) \Psi=-M\Psi,\quad \bar\Psi \cev D_\mu(A) \gamma_\mu=\bar\Psi M
\end{equation}
because we will have to deal with contact terms in the SymEFT as will become clear later.
Therefore $Q=\op\cup \opE$ contains both the on-shell basis $\op$ and a minimal basis of EOM-vanishing operators $\opE$ solely relevant in the presence of contact terms.

Similarly we can describe the lattice artifacts originating from the insertion of a discretised local field as
\begin{equation}
\Jop_\mathrm{eff}(x)=J(x)+a^{\nmin}\sum_i\nu_i\Jop_i^{(\nmin)}+\ord(a^{\nmin+1})
\end{equation}
with (bare) matching coefficients $\nu_i$.
The additional set of higher-dimensional operators of mass-dimension $([J]+\nmin)$ is constrained by the transformation properties of the local field on the lattice, here charge conjugation, parity, time reversal, and multiplication by a global flavour-dependent phase combined with any symmetry constraints from the action including the assumed flavour symmetries of the valence sector, discrete rotations etc.
We focus here on local quark bilinears of mass-dimension~3, i.e.,
\begin{equation}
J^{kl}(x)=[\bar{\psi}_k\Gamma\psi_l](x)\,,\quad \Gamma\in\{1,\gamma_5,\gamma_\mu,\gamma_\mu\gamma_5,i\sigma_{\mu\nu}\}.
\end{equation}
The notion of \emph{local} fields implies that insertions of those fields are physically separated when approaching the continuum limit, i.e., no contact terms arise in the lattice theory.
Except for the scalar bilinear, both choices of flavours $l=k$ and $l\neq k$ are being considered, while the scalar may only be non-singlet to avoid having to deal with mixing with the vacuum under renormalisation.

\subsection{Example: Pion decay constant}
To help to understand the general concepts, let us consider the pion decay constant $f_\pi$ extracted from the lattice
\begin{equation}
Z^{\hat{A}}\langle 0|\hat{A}_0^{ud}(x;a)|\pi(\vecn)\rangle=[m_\pi f_\pi](a),
\end{equation}
where the initial pion is at rest and $\hat{A}_0^{ud}$ is some discretisation of the temporal component of the axial vector -- potentially requiring multiplicative renormalisation with $Z^{\hat{A}}$.
Expanding the same quantity in our SymEFT then yields
\begin{align}
\frac{Z^{\hat{A}}\langle 0|\hat{A}_0^{ud}(x;a)|\pi(\vecn)\rangle}{m_\pi f_\pi}=1&+\frac{a^{\nmin}}{m_\pi f_\pi}\begin{pmatrix}
d^A \\[6pt]
-c^\Qop
\end{pmatrix}^\mathrm{T}
\begin{pmatrix}
Z^{A}     & 0 \\[6pt]
Z^{A\Qop} & Z^{\Qop}
\end{pmatrix}
\begin{pmatrix}
\langle 0|(A_0^{ud})^{(\nmin)}(x)|\pi(\vecn)\rangle \\[6pt]
\langle 0|A_0^{ud}(x)\tilde{\Qop}^{(\nmin)}(0)|\pi(\vecn)\rangle_\mathrm{c}
\end{pmatrix}\nonumber\\
&+\left(\substack{\text{contributions}\\\text{from }Z^{\hat{A}}}\right)+\ord(a^{\nmin+1})\,,\label{eq:mfpiNonDiag}
\end{align}
where $c^\Qop$ and $d^A$ are the renormalised matching coefficients matched at the renormalisation scale $\mu=1/a$ relevant for lattice artifacts, the tilde indicates the Fourier transform, and the matrix entries $Z^{A}$, $Z^{Q}$, $Z^{AQ}$ are themselves block-matrices acting on the different operator bases.
The subscript $\langle\ldots\rangle_\mathrm{c}$ denotes the connected contributions of the operator insertions from the SymEFT action.
Notice that we choose to perform the perturbative calculations of the SymEFT in dimensional regularisation in combination with the $\MSbar$ renormalisation scheme.
$A_0$ is Renormalisation Group Invariant (RGI).
Conceptually any renormalisation scheme can be used.
In principle the renormalisation of $\hat{A}_0^{ud}$ on the lattice may introduce more lattice artifacts, which has been indicated but will not be discussed further.
From the SymEFT point of view those contributions can be simply formulated as if the renormalisation condition contained in $Z^{\hat{A}}$ was yet another observable and $m_\pi f_\pi$ denotes the continuum quantity.
We assume that the asymptotic pion state does not introduce any further lattice artifacts as the artifacts associated to the interpolating fields employed on the lattice cancel out.
It should be apparent that the insertions of the basis for the SymEFT action will give rise to contact terms in eq.~\eqref{eq:mfpiNonDiag} when coinciding with the continuum local field $A_0$, which is the reason for the enlarged operator basis $\Qop$ and those contact terms get renormalised by $Z^{A\Qop}$.

Making a change of bases we may diagonalise all 1-loop mixing matrices\footnote{If the 1-loop mixing matrices turn out to be non-diagonalisable we can bring them into Jordan normal form, which in the following would give rise to additional overall powers in $\log(\gbar(1/a))$, see also the discussion of this more general case in \cite{Husung:2022kvi}.}
\begin{align}
\mu\frac{\rmd}{\rmd\mu}\begin{pmatrix}
(\mathcal{A}_0^{ud})^{(\nmin)}(x) \\[6pt]
A_0^{ud}(x)\tilde{\Qbase}^{(\nmin)}(0)
\end{pmatrix}_{\MSbar}&=
\begin{pmatrix}
T^{A}     & 0 \\[6pt]
T^{A\Qop} & T^{\Qop}
\end{pmatrix}
\mu\frac{\rmd}{\rmd\mu}
\begin{pmatrix}
(A_0^{ud})^{(\nmin)}(x) \\[6pt]
A_0^{ud}(x)\tilde{\Qop}^{(\nmin)}(0)
\end{pmatrix}_{\MSbar}\\[6pt]
&=-\frac{\gbar^2(\mu)}{(4\pi)^2}\begin{pmatrix}
(\gamma_0^{\mathcal{A}})^{(\nmin)}     & 0 \\[6pt]
0 & (\gamma_0^{\Qbase})^{(\nmin)}
\end{pmatrix}
\begin{pmatrix}
(\mathcal{A}_0^{ud})^{(\nmin)}(x) \\[6pt]
A_0^{ud}(x)\tilde{\Qbase}^{(\nmin)}(0)
\end{pmatrix}_{\MSbar}+\ord(\gbar^4)\,,\nonumber
\end{align}
where $\gamma_0^{\mathcal{A}}$ and $\gamma_0^{\Qbase}$ are again diagonal matrices acting on the different operator bases and $T^{A}$, $T^{A\Qop}$, $T^{\Qop}$ denote the appropriate changes of the operator bases.
With help of the Renormalisation Group we can then introduce RGI quantities \cite{Grunberg:1980ja}
\begin{equation}
X_{i;\mathrm{RGI}}^{(\nmin)}=\lim_{\mu\rightarrow\infty}[2b_0\gbar^2(\mu)]^{-(\gamma_0^X)_i^{(\nmin)}/(2b_0)}X_{i;\MSbar}^{(\nmin)}(\mu)\,,\quad X\in\{\mathcal{A},\Qbase\}\,.
\end{equation}
Meanwhile, performing the same change of bases on the (renormalised) matching coefficients yields
\begin{equation}
\begin{pmatrix}
\bar{d}^\mathcal{A} \\[6pt]
-\bar{c}^\Qbase
\end{pmatrix}^\mathrm{T}=
\begin{pmatrix}
d^A \\[6pt]
-c^\Qop
\end{pmatrix}^\mathrm{T}
\begin{pmatrix}
T^{A}     & 0 \\[6pt]
T^{A\Qop} & T^{\Qop}
\end{pmatrix}^{-1}
\end{equation}
for the matching coefficients of the diagonalised basis.
\emph{After} diagonalising the bases all contact divergences have been taken care of and we can finally make use of the freedom of adjusting the matching conditions.
This freedom allows us to set all matching coefficients of the EOM-vanishing extension to the basis of the SymEFT action to zero.\footnote{In principle this can be done iteratively to all orders in perturbation theory but here we are content with TL matching.}
We are only left with the diagonalised on-shell basis $\base$ instead of $\Qbase$, i.e., we impose
\begin{equation}
\begin{pmatrix}
\bar{d}^\mathcal{A} \\[6pt]
-\bar{c}^\Qbase
\end{pmatrix}\rightarrow\begin{pmatrix}
d^\mathcal{A} \\[6pt]
-c^\base \\[6pt]
-c^{\baseE}
\end{pmatrix}\stackrel{!}{=}\begin{pmatrix}
\hat{d}^{\mathcal{A}}[2b_0\gbar^2(1/a)]^{n^{\mathcal{A}}} \\[6pt]
-\hat{c}^{\base}[2b_0\gbar^2(1/a)]^{n^{\base}} \\[6pt]
0
\end{pmatrix}\times\left\{1+\ord(\gbar^2(1/a))\right\}\,.
\end{equation}
In the last equality we introduced $n_i^X\in\mathbb{Z}^+$ with $X\in\{\mathcal{A},\base\}$ to account for any vanishing tree-level  matching coefficients of our on-shell bases, in which case we assume $n_i^X=1$ as a conservative estimate.
If we did not carry the full basis $\Qop$ or $\Qbase$ respectively with us up to this point, $Z^{A\Qop}$ in eq.~\eqref{eq:mfpiNonDiag} and therefore $T^{A\Qop}$ would be insufficiently determined when working with an off-shell renormalisation prescription in the SymEFT. 
Evidently, a naive $a$-expansion is only sufficient to determine tree-level matching when combined with the final change of matching condition.

All of this combined lets us write the leading asymptotic lattice spacing dependence for the pion decay constant as
\begin{align}
\frac{Z^{\hat{A}}\langle 0|\hat{A}_0^{ud}(x;a)|\pi(\vecn)\rangle}{m_\pi f_\pi}=1&+a^{\nmin}\sum_i\hat{d}_i^{\mathcal{A}}[2b_0\gbar^2(1/a)]^{\hat{\Gamma}_i^{\mathcal{A}}}\frac{\langle 0|(\mathcal{A}_0^{ud})_{i;\mathrm{RGI}}^{(\nmin)}(x)|\pi(\vecn)\rangle}{m_\pi f_\pi}\nonumber\\
&-a^{\nmin}\sum_j\hat{c}_j^{\base}[2b_0\gbar^2(1/a)]^{\hat{\Gamma}_j^\base}\frac{\langle 0|A_0^{ud}(x)\tilde{\base}_{j;\mathrm{RGI}}^{(\nmin)}(0)|\pi(\vecn)\rangle_\mathrm{c}}{m_\pi f_\pi}\nonumber\\
&+\left(\substack{\text{contributions}\\\text{from }Z^{\hat{A}}}\right)+\ord\big(a^{\nmin+1},a^{\nmin}\gbar^{2\hat{\Gamma}+2}(1/a)\big)\,,\label{eq:mfpiAsymp}
\end{align}
where we denote the leading powers in $\gbar^2(1/a)$ after matching as
\begin{equation}
\hat{\Gamma}_i^{X}=\frac{(\gamma_0^{X})_i^{(\nmin)}}{2b_0}+n_i^{X},\quad X\in\{\mathcal{A},\base\}.
\end{equation}
In the more general case of a continuum local field without vanishing 1-loop anomalous dimension, i.e., $\gamma_0^{\Jop}\neq 0$ we would need to adjust this to the difference
\begin{equation}
\hat{\Gamma}_i^{\Jbase}=\hat{\gamma}_i^{\Jbase}+n_i^{\Jbase}\,,\quad \hat{\gamma}_i^{\Jbase}=\frac{(\gamma_0^{\Jbase})_i^{(\nmin)}-\gamma_0^{\Jop}}{2b_0}\,,
\end{equation}
where $\Jbase$ is the proper diagonalised higher-dimensional operator basis analogously to $\mathcal{A}$.
In eq.~\eqref{eq:mfpiAsymp} all lattice-spacing dependence is now absorbed into the prefactors, while the remaining RGI matrix elements are unknown constants.

To conclude our initial example, we also need to specify what minimal basis to use.
For the axial-vector, we have the minimal on-shell basis at $\ord(a)$ in agreement with~\cite{Luscher:1996sc,Bhattacharya:2003nd,Bhattacharya:2005rb}
\newcommand{\cevvec}[1]{\overset{\longleftrightarrow}{#1}}
\def\mathrlap#1{#1}
\begin{align}
\A^{(1)}_1&=\partial_\mu\mathrm{P}^{kl},&
\A^{(1)}_2&=\frac{m_{k+l}}{2}\mathrm{A}_\mu^{kl},&
\A^{(1)}_3&=\tr(M)\mathrm{A}_\mu^{kl},
\end{align}
where $m_{k\pm l}=\frac{m_k\pm m_l}{2}$.
Similarly, we find at $\ord(a^2)$
\begin{align}
\A^{(2)}_1&=\delta_{\mu\rho\lambda}\bar{q}_k\gamma_\rho\gamma_5\cevvec{D^{\mathrlap{\smash{2}}}_{\lambda}}q_l,&
\A^{(2)}_2&=\bar{q}_k\gamma_\rho \tilde{F}_{\rho\mu}q_l,\nonumber\\
\A^{(2)}_3&=m_{k-l}\bar{q}_k(\cev{D}_\mu -D_\mu)\gamma_5q_l,\vphantom{\frac{\delta_{kl}}{g_0^2}}&
\A^{(2)}_{4}&=\frac{\delta_{kl}\delta_{\mu\nu\rho\sigma}}{g_0^2}\tr(D_\nu F_{\rho\lambda}\tilde{F}_{\sigma\lambda}),\nonumber\\
\A^{(2)}_{5}&=\frac{\delta_{kl}}{g_0^2}\tr(D_\rho F_{\rho\lambda}\tilde{F}_{\mu\lambda}),&
\A^{(2)}_6&=\delta_{\mu\rho\lambda}\partial_\rho^2\mathrm{A}_\lambda^{kl},\nonumber\\
\A^{(2)}_7&=\partial^2\mathrm{A}_\mu^{kl},&
\A^{(2)}_8&=\frac{m_k^2+m_l^2}{2}\mathrm{A}_\mu^{kl},\nonumber\\
\A^{(2)}_{9}&=\frac{\delta_{kl}}{g_0^2}\partial_\mu\tr(F_{\nu\rho}\tilde F_{\nu\rho}),&
\A^{(2)}_{9+(j<3)}&=\frac{m_{k+l}}{2}\A^{(1)}_j,\nonumber\\
\A^{(2)}_{11+j}&=\tr(M)\A^{(1)}_j,&
\A^{(2)}_{15}&=\tr(M^2)\mathrm{A}_\mu^{kl}.
\end{align}
From 1-loop renormalisation of these bases we find the corresponding $\hat{\gamma}_i^{\mathcal{A}}$, which are listed in table~\ref{tab:1loopAD} for $\Nf=2,3$.
Table~\ref{tab:1loopAD} also includes the results for all the other local fermion bilinears mentioned in the beginning but not discussed here in detail.
\begin{table}
\caption{(Distinct) 1-loop anomalous dimensions found for the local bilinears at $\Nf=2,3$ flavours in 3-colour lattice QCD rounded to the third decimal.
Keep in mind that $(\hat{\gamma}^\Jbase+1)$ are not added but will arise from loop-suppressed contributions.
\uwave{Underwiggled} numbers occur only for flavour-singlets and therefore may only contribute if $k=l$.
\underline{Underlined} numbers belong to massive contributions.
\dotuline{Underdotted} numbers correspond to massive contributions, that vanish both in the mass-degenerate case and for $k=l$.
}\label{tab:1loopAD}\centering
\begin{tabular}{lc|l|l}
  & $\Nf$ & $\big(\hat{\gamma}^\Jbase\big)^{(1)}$ & $\big(\hat{\gamma}^\Jbase\big)^{(2)}$ \\\hline
scalar ($k\neq l$) & 2     &  \underline{$0.414$} & $0$, $0.483$, \underline{$0.828$} \\
                        & 3     &  \underline{$0.444$} & $0$, $0.519$, \underline{$0.889$} \\
pseudo-scalar           & 2     &  \uwave{$-0.586$}, \underline{$0.414$} & \underline{\uwave{$-0.172$}}, $0$, $0.483$, \underline{$0.828$} \\
                        & 3     &  \uwave{$-0.556$}, \underline{$0.444$} & \underline{\uwave{$-0.111$}}, $0$, $0.519$, \underline{$0.889$} \\
vector                  & 2     &  $0.138$, \underline{$0.414$} & $0$, $0.368$, \underline{$0.552$}, $0.575$, \underline{$0.828$} \\
                        & 3     &  $0.148$, \underline{$0.444$} & $0$, $0.395$, \underline{$0.593$}, $0.617$, \underline{$0.889$} \\
axial-vector            & 2     &  $-0.414$, \underline{$0.414$} & \uwave{$-1$}, $0$, $0.368$, \uwave{$0.506$}, \dotuline{$0.552$}, \uwave{$0.559$}, $0.575$, \underline{$0.828$},  \uwave{$1.085$} \\
                        & 3     &  $-0.444$, \underline{$0.444$} & \uwave{$-1$}, $0$, $0.395$, \dotuline{$0.593$}, \uwave{$0.595$}, $0.617$, \underline{$0.889$}, \uwave{$1.244$} \\
tensor                  & 2     &  $-0.138$, \underline{$0.414$} & $0$, \underline{$0.276$}, $0.46$, $0.563$, $0.69$, \underline{$0.828$} \\
                        & 3     &  $-0.148$, \underline{$0.444$} & $0$, \underline{$0.296$}, $0.494$, $0.605$, $0.741$, \underline{$0.889$}
\end{tabular}
\end{table}
If we were only interested in those (minimal) powers we could stop at this point and ignore any potential suppression from vanishing tree-level matching coefficients.
Otherwise also contact interactions must be taken into account, but those details are beyond the scope of this proceedings contribution.
The full bases for the other fermion bilinears and the impact of contact interactions will be made available \cite{Husung:inprep}.

\subsection{\boldmath Negative powers of $\hat{\gamma}^\Jbase$}
Rather than going in detail through all the other fermion bilinears mentioned, we will now pick out particular powers $\hat{\gamma}^{\Jbase}$ -- the negative ones from table~\ref{tab:1loopAD} -- and trace back where those originate from and whether we can do something about them.
Fortunately there is only a very limited number of those negative powers up to $\ord(a^2)$.
\paragraph{Pseudo-scalar} Only for the flavour-singlet case we find a severely negative power at $\ord(a)$ due to
\begin{equation}
\Pbase^{(1)}\sim\frac{\delta_{kl}}{g_0^2}\tr(F_{\mu\nu}\tilde{F}_{\mu\nu})\,,
\end{equation}
where $\sim$ denotes the operator in the non-diagonal basis eventually giving rise to the power in the coupling being discussed here after the full diagonalisation of bases and contact-term subtractions.\footnote{It is non-trivial that such a mapping exists to this order in the lattice spacing that leaves this diagonal entry of the 1-loop anomalous dimension matrix in its place.}
This term may only arise for a theory without exact lattice chiral symmetry, i.e., here it only plays a role for Wilson quarks.
In a theory without an $\ord(a)$ improved lattice action, the contact term between the continuum field and the operator $\frac{i}{4}\bar{\Psi}\sigma_{\mu\nu}F_{\mu\nu}\Psi$ from the SymEFT action will give rise to a non-vanishing tree-level matching coefficient of this term in the diagonalised basis.
Otherwise, the matching coefficient will usually be loop-suppressed.

In the massless limit and finite volume, these contributions will be subject to automatic $\ord(a)$ improvement by a discrete 2-flavour continuum symmetry
\begin{equation}
\bar{\Psi}\rightarrow i\bar{\Psi}\gamma_5\tau^3,\quad \Psi\rightarrow i\gamma_5\tau^3\Psi,
\end{equation}
under which this $\ord(a)$ term transforms with opposite sign than the continuum field, similarly to twisted mass QCD~\cite{Aoki:2006gh,Sint:2007ug}.
Consequently, those contributions will only affect $\ord(a^2)$ by the interplay with other $\ord(a)$ terms cancelling the relative sign-change under this transformation either from other local fields or the SymEFT action.
The latter will also yield contact terms at $\ord(a^2)$ that have not been considered here but those will only affect the matching of the mass-dimension~5 operator basis.
Notice that the asymptotic lattice spacing dependence of different $\ord(a)$ terms combined will also give rise to $\ord(a^2)$ effects, because  one finds for any two $\ord(a)$ terms 
\begin{equation}
a[2b_0\gbar^2(1/a)]^{\hat{\gamma}_1}\times a[2b_0\gbar^2(1/a)]^{\hat{\gamma}_2}=a^2[2b_0\gbar^2(1/a)]^{\hat{\gamma}_1+\hat{\gamma}_2}.
\end{equation}

The same term multiplied by a quark mass reappears at $\ord(a^2)$, i.e., this additional contribution only plays a role for non-vanishing quark masses.
Again it yields a negative power $\hat{\gamma}^{\Jbase}$ but much closer to zero.

\paragraph{Axial-vector} At $\ord(a)$ the operator 
\begin{equation}
\Abase^{(1)}\sim\A^{(1)}_1
\end{equation}
gives rise to severely negative powers.
It is not allowed to contribute for lattice actions with exact lattice chiral symmetry.
The term $\Abase^{(1)}$ is expected to contribute at tree-level for non-strictly-local lattice bilinears, e.g., unimproved conserved currents.
Again, the remarks from the pseudo-scalar about automatic $\ord(a)$ improvement in the massless finite volume theory apply.

In case of an explicitly\footnote{When relying on automatic $\ord(a)$ improvement the terms discussed before will be relevant.} $\ord(a)$ improved theory, we only find for the flavour singlet a severely negative power at $\ord(a^2)$ due to the term
\begin{equation}
\Abase^{(2)}\sim\frac{\delta_{kl}}{g_0^2}\partial_\mu\tr(F_{\nu\rho}\tilde F_{\nu\rho})\,.
\end{equation}
This term will arise already at tree-level matching due to contact interactions with operators from the SymEFT action such as $\bar{\Psi}\gamma_\mu D_\mu^3\Psi$.
Its impact could be reduced by (at least) Symanzik tree-level $\ord(a^2)$ improvement of the lattice action as suggested, e.g., in \cite{Husung:2022kvi}.

\paragraph{Tensor} Only at $\ord(a)$ we find a slightly negative power due to
\begin{equation}
\Tbase^{(1)}\sim \partial_\mu V_\nu-\partial_\nu V_\mu\,.
\end{equation}
This term is again expected to arise at tree-level matching for non-strictly-local lattice bilinears but also arises from the contact interaction between the continuum field and the operator $\frac{i}{4}\bar{\Psi}\sigma_{\mu\nu}F_{\mu\nu}\Psi$ from the SymEFT action.
As was the case for the pseudo-scalar and axial-vector, this term is only allowed for lattice actions without exact lattice chiral symmetry and also the statements on automatic $\ord(a)$ improvement in the massless finite volume theory apply.

\section{Conclusion}
After this first glance at local fermion bilinears of mass-dimension~3 we have a complete overview of the asymptotically leading lattice artifacts of the form $a^{\nmin}[2b_0\gbar^2(1/a)]^{\hat{\gamma}_i^{\Jbase}}$ without taking potential suppression from tree-level matching into account.
The resulting picture is a bit more mixed than for the lattice actions discussed so far~\cite{Husung:2019ytz,Husung:2022kvi}.
In particular at $\ord(a)$ and flavour-singlet $\ord(a^2)$ we find some anomalous dimensions that may slow down the convergence to the continuum limit significantly and should be dealt with.
Here this can already be achieved by simple perturbative (tree-level) Symanzik improvement shifting the problematic powers beyond classical $a^{\nmin}$ corrections.
In practice, $\ord(a)$ improvement of the local fields is frequently considered in the literature, see e.g.~\cite{Luscher:1996sc,Capitani:1999ay,Capitani:2000xi,Bhattacharya:2003nd,Bhattacharya:2005rb,Heitger:2020zaq}.
The results also highlight once more the benefits of explicit Symanzik $\ord(a)$ improvement of Wilson quarks~\cite{Luscher:1996sc} over automatic $\ord(a)$ improvement.
The latter only postpones the problematic impact of negative powers to $\ord(a^2)$ and further introduces additional contact terms in the SymEFT description affecting tree-level matching at $\ord(a^2)$.
Those contact terms do not introduce too problematic powers in the coupling for non-singlet currents.
As an example, the interplay of the unimproved massless (or twisted) Wilson quark action at $\Nf=2$ with the local axial-vector will introduce in eq.~\eqref{eq:mfpiAsymp} a non-vanishing contribution of the asymptotic form $a^2[2b_0\gbar^2(1/a)]^{-0.276}$.
This is still far from $\hat{\Gamma}\simeq -3$ but could be avoided easily.

In an explicitly $\ord(a)$ improved setup, the leading lattice artifacts originating from the non-singlet bilinears will at most be classical $a^2$ or converge even faster.

Overall, these additional powers $\hat{\gamma}^{\Jbase}$ complement the ones already found for the lattice action and should be incorporated into continuum extrapolations by varying the ansatz beyond naive integer-powers in the lattice spacing to $a^{\nmin}[2b_0\gbar(1/a)]^{\hat\Gamma}$ with $\hat{\Gamma}$ varied in the range of powers suggested by leading-order SymEFT calculations.
A caveat remains whether nowadays lattice simulations are at sufficiently small lattice spacings such that leading-order SymEFT is applicable or in other words the running coupling $\alpha_{\MSbar}(1/a)=\gbar^2(1/a)/(4\pi)$ is sufficiently small.
Of course higher powers in the lattice spacing should be considered as well to have a reasonably conservative estimate of this systematic uncertainty.
The fact that the range of (fine) lattice spacings typically available is insufficient to distinguish these different powers is not an argument against those variations but, on the contrary, shows the lack of control over this systematic error.
Finer lattice spacings would therefore always be beneficial but are too expensive at the moment except for step-scaling studies in small volume such as~\cite{DallaBrida:2019wur}.

The results presented here are valid only for \emph{on-shell} matrix-elements of \emph{local} fields.
For integrated correlation functions such as the integrated vector-vector 2-point function required for predictions of the QCD contributions to the anomalous magnetic moment of the muon~\cite{Aoyama:2020ynm,Borsanyi:2020mff,Kuberski:2023qgx}, additional contact terms of the bilinears arise even on the lattice that have not been considered here.
Despite this limitation, the results presented here should be sufficient for so called \emph{window quantities}~\cite{Aubin:2022hgm,FermilabLattice:2022izv,Ce:2022kxy,RBC:2023pvn,Kuberski:2023qgx} at intermediate or large distances where those contact terms should be sufficiently suppressed.
The use of off-shell renormalisation schemes such as RI/(S)MOM~\cite{Martinelli:1994ty,Sturm:2009kb} is not being covered as this would require use of a significantly enlarged operator basis constrained by BRST invariance rather than gauge invariance for both the action and the local fields.
Furthermore, off-shell renormalisation conditions invalidate the use of the EOMs even beyond contact terms and introduce gauge-choice dependence.
In contrast, on-shell renormalisation schemes are covered but still may require the inclusion of additional lattice artifacts from the renormalisation condition as indicated in eq.~\eqref{eq:mfpiAsymp}.
An example of such an on-shell renormalisation scheme would be the Schr\"odinger functional scheme~\cite{Jansen:1995ck} for which the additional lattice artifacts from the time-boundary have already been discussed to $\ord(a)$ in SymEFT~\cite{Husung:2019ytz}.

The scripts used to perform the necessary calculations are publicly available\footnote{\url{https://github.com/nikolai-husung/Symanzik-QCD-workflow}} including some automation for the 1-loop renormalisation.
Finding the minimal bases is not automated and will require quite some work for each new local field being considered.\\

\noindent\textbf{Acknowledgements.} I am grateful to Rainer Sommer, Chris Sachrajda, and Jonathan Flynn for discussions on the automatic $\ord(a)$ improvement of the massless theory in finite volume.
I thank Gregorio Herdoiza for helpful suggestions on the manuscript.
The author acknowledges funding by the STFC consolidated grant ST/T000775/1.

\bibliographystyle{JHEP}
\bibliography{lat23.bbl}

\providecommand{\href}[2]{#2}\begingroup\raggedright\begin{thebibliography}{10}

\bibitem{Balog:2009yj}
J.~Balog, F.~Niedermayer and P.~Weisz, \emph{{Logarithmic corrections to
  O($a^2$) lattice artifacts}},
  \href{https://doi.org/10.1016/j.physletb.2009.04.082}{\emph{Phys. Lett.}
  {\bfseries B676} (2009) 188}
  [\href{https://arxiv.org/abs/0901.4033}{{\ttfamily 0901.4033}}].

\bibitem{Balog:2009np}
J.~Balog, F.~Niedermayer and P.~Weisz, \emph{{The Puzzle of apparent linear
  lattice artifacts in the 2d non-linear sigma-model and Symanzik's solution}},
  \href{https://doi.org/10.1016/j.nuclphysb.2009.09.007}{\emph{Nucl. Phys.}
  {\bfseries B824} (2010) 563}
  [\href{https://arxiv.org/abs/0905.1730}{{\ttfamily 0905.1730}}].

\bibitem{Pasztor:2019rqu}
{\scshape ATLAS, CMS} collaboration, \emph{{Precision tests of the Standard
  Model at the LHC with the ATLAS and CMS detectors}},
  \href{https://doi.org/10.22323/1.353.0005}{\emph{PoS} {\bfseries FFK2019}
  (2020) 005}.

\bibitem{LHCb:2008vvz}
{\scshape LHCb} collaboration, \emph{{The LHCb Detector at the LHC}},
  \href{https://doi.org/10.1088/1748-0221/3/08/S08005}{\emph{JINST} {\bfseries
  3} (2008) S08005}.

\bibitem{Belle-II:2018jsg}
{\scshape Belle-II} collaboration, \emph{{The Belle II Physics Book}},
  \href{https://doi.org/10.1093/ptep/ptz106}{\emph{PTEP} {\bfseries 2019}
  (2019) 123C01} [\href{https://arxiv.org/abs/1808.10567}{{\ttfamily
  1808.10567}}].

\bibitem{Muong-2:2023cdq}
{\scshape Muon g-2} collaboration, \emph{{Measurement of the Positive Muon
  Anomalous Magnetic Moment to 0.20~ppm}},
  \href{https://doi.org/10.1103/PhysRevLett.131.161802}{\emph{Phys. Rev. Lett.}
  {\bfseries 131} (2023) 161802}
  [\href{https://arxiv.org/abs/2308.06230}{{\ttfamily 2308.06230}}].

\bibitem{Husung:2019ytz}
N.~Husung, P.~Marquard and R.~Sommer, \emph{{Asymptotic behavior of cutoff
  effects in Yang-Mills theory and in Wilson's lattice QCD}},
  \href{https://doi.org/10.1140/epjc/s10052-020-7685-4}{\emph{Eur.\ Phys.\ J.\
  C} {\bfseries 80} (2020) 200}
  [\href{https://arxiv.org/abs/1912.08498}{{\ttfamily 1912.08498}}].

\bibitem{Husung:2021mfl}
N.~Husung, P.~Marquard and R.~Sommer, \emph{{The asymptotic approach to the
  continuum of lattice QCD spectral observables}},
  \href{https://doi.org/10.1016/j.physletb.2022.137069}{\emph{Phys. Lett. B}
  {\bfseries 829} (2022) 137069}
  [\href{https://arxiv.org/abs/2111.02347}{{\ttfamily 2111.02347}}].

\bibitem{Husung:2022kvi}
N.~Husung, \emph{{Logarithmic corrections to O(a) and O($a^2$) effects in
  lattice QCD with Wilson or Ginsparg\textendash{}Wilson quarks}},
  \href{https://doi.org/10.1140/epjc/s10052-023-11258-8}{\emph{Eur. Phys. J. C}
  {\bfseries 83} (2023) 142}
  [\href{https://arxiv.org/abs/2206.03536}{{\ttfamily 2206.03536}}].

\bibitem{FlavourLatticeAveragingGroupFLAG:2021npn}
{\scshape Flavour Lattice Averaging Group (FLAG)} collaboration, \emph{{FLAG
  Review 2021}},
  \href{https://doi.org/10.1140/epjc/s10052-022-10536-1}{\emph{Eur. Phys. J. C}
  {\bfseries 82} (2022) 869}
  [\href{https://arxiv.org/abs/2111.09849}{{\ttfamily 2111.09849}}].

\bibitem{Symanzik:1979ph}
K.~Symanzik, \emph{{Cutoff dependence in lattice $\phi_4^4$ theory}},
  \href{https://doi.org/10.1007/978-1-4684-7571-5_18}{\emph{NATO Sci. Ser. B}
  {\bfseries 59} (1980) 313}.

\bibitem{Symanzik:1981hc}
K.~Symanzik, \emph{{Some Topics in Quantum Field Theory}},  in
  \emph{{Mathematical Problems in Theoretical Physics. Proceedings, 6th
  International Conference on Mathematical Physics, West Berlin, Germany,
  August 11-20, 1981}}, pp.~47--58, 1981.

\bibitem{Symanzik:1983dc}
K.~Symanzik, \emph{{Continuum Limit and Improved Action in Lattice Theories. 1.
  Principles and $\phi^4$ Theory}},
  \href{https://doi.org/10.1016/0550-3213(83)90468-6}{\emph{Nucl. Phys.}
  {\bfseries B226} (1983) 187}.

\bibitem{Symanzik:1983gh}
K.~Symanzik, \emph{{Continuum Limit and Improved Action in Lattice Theories. 2.
  O(N) Nonlinear Sigma Model in Perturbation Theory}},
  \href{https://doi.org/10.1016/0550-3213(83)90469-8}{\emph{Nucl. Phys.}
  {\bfseries B226} (1983) 205}.

\bibitem{Bar:2005tu}
O.~{B\"ar}, C.~Bernard, G.~Rupak and N.~Shoresh, \emph{{Chiral perturbation
  theory for staggered sea quarks and Ginsparg-Wilson valence quarks}},
  \href{https://doi.org/10.1103/PhysRevD.72.054502}{\emph{Phys. Rev. D}
  {\bfseries 72} (2005) 054502}
  [\href{https://arxiv.org/abs/hep-lat/0503009}{{\ttfamily hep-lat/0503009}}].

\bibitem{Wilson:1974}
K.~G. Wilson, \emph{Confinement of quarks},
  \href{https://doi.org/10.1103/PhysRevD.10.2445}{\emph{Phys. Rev. D}
  {\bfseries 10} (1974) 2445}.

\bibitem{Wilson:1975id}
K.~G. Wilson, \emph{{Quarks and Strings on a Lattice}},  in \emph{{New
  Phenomena in Subnuclear Physics: Proceedings, International School of
  Subnuclear Physics, Erice, Sicily, Jul 11-Aug 1 1975. Part A}}, p.~99, 1975.

\bibitem{Ginsparg:1981bj}
P.~H. Ginsparg and K.~G. Wilson, \emph{{A Remnant of Chiral Symmetry on the
  Lattice}}, \href{https://doi.org/10.1103/PhysRevD.25.2649}{\emph{Phys. Rev.}
  {\bfseries D25} (1982) 2649}.

\bibitem{Grunberg:1980ja}
G.~Grunberg, \emph{{Renormalization Group Improved Perturbative QCD}},
  \href{https://doi.org/10.1016/0370-2693(80)90402-5}{\emph{Phys. Lett.}
  {\bfseries 95B} (1980) 70}.

\bibitem{Luscher:1996sc}
M.~{L\"uscher}, S.~Sint, R.~Sommer and P.~Weisz, \emph{{Chiral symmetry and
  O(a) improvement in lattice QCD}},
  \href{https://doi.org/10.1016/0550-3213(96)00378-1}{\emph{Nucl. Phys.}
  {\bfseries B478} (1996) 365}
  [\href{https://arxiv.org/abs/hep-lat/9605038}{{\ttfamily hep-lat/9605038}}].

\bibitem{Bhattacharya:2003nd}
T.~Bhattacharya, R.~Gupta, W.-j. Lee, S.~R. Sharpe and J.~M. Wu,
  \emph{{Improved bilinears in unquenched lattice QCD}},
  \href{https://doi.org/10.1016/S0920-5632(03)02608-2}{\emph{Nucl. Phys. B
  Proc. Suppl.} {\bfseries 129} (2004) 441}
  [\href{https://arxiv.org/abs/hep-lat/0309087}{{\ttfamily hep-lat/0309087}}].

\bibitem{Bhattacharya:2005rb}
T.~Bhattacharya, R.~Gupta, W.~Lee, S.~R. Sharpe and J.~M. Wu, \emph{{Improved
  bilinears in lattice QCD with non-degenerate quarks}},
  \href{https://doi.org/10.1103/PhysRevD.73.034504}{\emph{Phys. Rev. D}
  {\bfseries 73} (2006) 034504}
  [\href{https://arxiv.org/abs/hep-lat/0511014}{{\ttfamily hep-lat/0511014}}].

\bibitem{Husung:inprep}
N.~Husung, \emph{{Lattice artifacts of local fermion bilinears up to
  O($a^2$)}}, {\emph{in preparation}}.

\bibitem{Aoki:2006gh}
S.~Aoki and O.~{B\"ar}, \emph{{Automatic O(a) improvement for twisted-mass
  QCD}}, \href{https://doi.org/10.22323/1.032.0165}{\emph{PoS} {\bfseries
  LAT2006} (2006) 165} [\href{https://arxiv.org/abs/hep-lat/0610098}{{\ttfamily
  hep-lat/0610098}}].

\bibitem{Sint:2007ug}
S.~Sint, \emph{{Lattice QCD with a chiral twist}},  in \emph{{Workshop on
  Perspectives in Lattice QCD Nara, Japan, October 31-November 11, 2005}},
  2007, \href{https://arxiv.org/abs/hep-lat/0702008}{{\ttfamily
  hep-lat/0702008}}, \href{https://doi.org/10.1142/9789812790927_0004}{DOI}.

\bibitem{Capitani:1999ay}
S.~Capitani, M.~{G\"ockeler}, R.~Horsley, P.~E.~L. Rakow and G.~Schierholz,
  \emph{{On-shell and off-shell improvement for Ginsparg-Wilson fermions}},
  \href{https://doi.org/10.1016/S0920-5632(00)91837-1}{\emph{Nucl. Phys. B
  Proc. Suppl.} {\bfseries 83} (2000) 893}
  [\href{https://arxiv.org/abs/hep-lat/9909167}{{\ttfamily hep-lat/9909167}}].

\bibitem{Capitani:2000xi}
S.~Capitani, M.~{G\"ockeler}, R.~Horsley, H.~Perlt, P.~E.~L. Rakow,
  G.~Schierholz et~al., \emph{{Renormalization and off-shell improvement in
  lattice perturbation theory}},
  \href{https://doi.org/10.1016/S0550-3213(00)00590-3}{\emph{Nucl. Phys. B}
  {\bfseries 593} (2001) 183}
  [\href{https://arxiv.org/abs/hep-lat/0007004}{{\ttfamily hep-lat/0007004}}].

\bibitem{Heitger:2020zaq}
{\scshape ALPHA} collaboration, \emph{{The renormalised $\mathrm{O}(a)$
  improved vector current in three-flavour lattice QCD with Wilson quarks}},
  \href{https://doi.org/10.1140/epjc/s10052-021-09037-4}{\emph{Eur. Phys. J. C}
  {\bfseries 81} (2021) 254}
  [\href{https://arxiv.org/abs/2010.09539}{{\ttfamily 2010.09539}}].

\bibitem{DallaBrida:2019wur}
M.~Dalla~Brida and A.~Ramos, \emph{{The gradient flow coupling at high-energy
  and the scale of SU(3) Yang--Mills theory}},
  \href{https://doi.org/10.1140/epjc/s10052-019-7228-z}{\emph{Eur. Phys. J. C}
  {\bfseries 79} (2019) 720}
  [\href{https://arxiv.org/abs/1905.05147}{{\ttfamily 1905.05147}}].

\bibitem{Aoyama:2020ynm}
T.~Aoyama et~al., \emph{{The anomalous magnetic moment of the muon in the
  Standard Model}},
  \href{https://doi.org/10.1016/j.physrep.2020.07.006}{\emph{Phys. Rept.}
  {\bfseries 887} (2020) 1} [\href{https://arxiv.org/abs/2006.04822}{{\ttfamily
  2006.04822}}].

\bibitem{Borsanyi:2020mff}
S.~Borsanyi et~al., \emph{{Leading hadronic contribution to the muon magnetic
  moment from lattice QCD}},
  \href{https://doi.org/10.1038/s41586-021-03418-1}{\emph{Nature} {\bfseries
  593} (2021) 51} [\href{https://arxiv.org/abs/2002.12347}{{\ttfamily
  2002.12347}}].

\bibitem{Kuberski:2023qgx}
S.~Kuberski, \emph{{Muon $g-2$: Lattice calculations of the hadronic vacuum
  polarization}},  12, 2023, \href{https://arxiv.org/abs/2312.13753}{{\ttfamily
  2312.13753}}.

\bibitem{Aubin:2022hgm}
C.~Aubin, T.~Blum, M.~Golterman and S.~Peris, \emph{{Muon anomalous magnetic
  moment with staggered fermions: Is the lattice spacing small enough?}},
  \href{https://doi.org/10.1103/PhysRevD.106.054503}{\emph{Phys. Rev. D}
  {\bfseries 106} (2022) 054503}
  [\href{https://arxiv.org/abs/2204.12256}{{\ttfamily 2204.12256}}].

\bibitem{FermilabLattice:2022izv}
{\scshape Fermilab Lattice, MILC, HPQCD} collaboration, \emph{{Windows on the
  hadronic vacuum polarization contribution to the muon anomalous magnetic
  moment}}, \href{https://doi.org/10.1103/PhysRevD.106.074509}{\emph{Phys. Rev.
  D} {\bfseries 106} (2022) 074509}
  [\href{https://arxiv.org/abs/2207.04765}{{\ttfamily 2207.04765}}].

\bibitem{Ce:2022kxy}
M.~C\`e et~al., \emph{{Window observable for the hadronic vacuum polarization
  contribution to the muon g-2 from lattice QCD}},
  \href{https://doi.org/10.1103/PhysRevD.106.114502}{\emph{Phys. Rev. D}
  {\bfseries 106} (2022) 114502}
  [\href{https://arxiv.org/abs/2206.06582}{{\ttfamily 2206.06582}}].

\bibitem{RBC:2023pvn}
{\scshape RBC, UKQCD} collaboration, \emph{{Update of Euclidean windows of the
  hadronic vacuum polarization}},
  \href{https://doi.org/10.1103/PhysRevD.108.054507}{\emph{Phys. Rev. D}
  {\bfseries 108} (2023) 054507}
  [\href{https://arxiv.org/abs/2301.08696}{{\ttfamily 2301.08696}}].

\bibitem{Martinelli:1994ty}
G.~Martinelli, C.~Pittori, C.~T. Sachrajda, M.~Testa and A.~Vladikas, \emph{{A
  General method for nonperturbative renormalization of lattice operators}},
  \href{https://doi.org/10.1016/0550-3213(95)00126-D}{\emph{Nucl. Phys. B}
  {\bfseries 445} (1995) 81}
  [\href{https://arxiv.org/abs/hep-lat/9411010}{{\ttfamily hep-lat/9411010}}].

\bibitem{Sturm:2009kb}
C.~Sturm, Y.~Aoki, N.~H. Christ, T.~Izubuchi, C.~T.~C. Sachrajda and A.~Soni,
  \emph{{Renormalization of quark bilinear operators in a momentum-subtraction
  scheme with a nonexceptional subtraction point}},
  \href{https://doi.org/10.1103/PhysRevD.80.014501}{\emph{Phys. Rev. D}
  {\bfseries 80} (2009) 014501}
  [\href{https://arxiv.org/abs/0901.2599}{{\ttfamily 0901.2599}}].

\bibitem{Jansen:1995ck}
K.~Jansen, C.~Liu, M.~{L\"uscher}, H.~Simma, S.~Sint, R.~Sommer et~al.,
  \emph{{Nonperturbative renormalization of lattice QCD at all scales}},
  \href{https://doi.org/10.1016/0370-2693(96)00075-5}{\emph{Phys. Lett. B}
  {\bfseries 372} (1996) 275}
  [\href{https://arxiv.org/abs/hep-lat/9512009}{{\ttfamily hep-lat/9512009}}].

\end{thebibliography}\endgroup

\end{document}